\documentclass[twocolumn,aps,prb,amsmath,amssymb]{revtex4}

\usepackage{graphicx}
\usepackage{dcolumn}
\usepackage{bm}

\usepackage{hyperref}

\newcommand{\ket}[1]{\ensuremath{|#1\rangle}}

\begin{document}

\title{Competition between electric field and magnetic field noise\\
 in the decoherence of a single spin in diamond}
\author{P. Jamonneau$^{1}$}
\author{M. Lesik$^{1}$}
\author{J. P. Tetienne$^{1}$}
\author{I. Alvizu$^{2}$}
\author{L. Mayer$^{1}$}
\author{A. Dr\'eau$^{1}$}
\author{S. Kosen$^{1}$}
\author{J.-F. Roch$^{1}$}
\author{S. Pezzagna$^{3}$}
\author{J. Meijer$^{3}$}
\author{T. Teraji$^{4}$}
\author{Y. Kubo$^{5}$}
\author{P. Bertet$^{5}$}
\author{J. R. Maze$^{2}$}
\author{V. Jacques$^{1,6}$}
\email{vincent.jacques@umontpellier.fr}
\affiliation{$^{1}$Laboratoire Aim\'{e} Cotton, CNRS, Universit\'{e} Paris-Sud and Ecole Normale Sup\'erieure de Cachan, 91405 Orsay, France}
\affiliation{$^{2}$Facultad de F\'{i}sica, Pontificia Universidad Cat\'{o}lica de Chile, Santiago 7820436, Chile}
\affiliation{$^{3}$ Department of Nuclear Solid State Physics, Institute for Experimental Physics II, Universitat Leipzig, 
LinnŽstr. 5, 04103 Leipzig, Germany}
\affiliation{$^{4}$ National Institute for Materials Science, 1-1 Namiki, Tsukuba, Ibaraki 305-0044, Japan}
\affiliation{$^{5}$ Quantronics group, Service de Physique de l'Etat Condens\'e, DSM/IRAMIS/SPEC, CNRS UMR 3680, CEA Saclay, 91191 Gif-sur-Yvette, France}
\affiliation{$^{6}$ Laboratoire Charles Coulomb, Universit\'{e} de Montpellier and CNRS, 34095 Montpellier, France}

\begin{abstract}
We analyze the impact of electric field and magnetic field fluctuations in the decoherence of the electronic spin associated with a single nitrogen-vacancy (NV) defect in diamond by engineering spin eigenstates protected either against magnetic noise or against electric noise. The competition between these noise sources is analyzed quantitatively by changing their relative strength through modifications of the environment. This study provides significant insights into the decoherence of the NV electronic spin, which is valuable for quantum metrology and sensing applications.
\end{abstract}

\maketitle

Improving the coherence time of solid-state spin qubits is a central challenge in quantum technologies. Decoherence is induced by fluctuations of the local environment and can be mitigated by following several strategies. On one hand, the tools of material science can be exploited to engineer host samples with quantum grade purity~\cite{Balasu2009}. As an example, millisecond-long coherence times have been achieved for electron spin impurities in isotopically purified diamond samples at room temperature~\cite{Balasu2009,Mizuochi}, while few seconds can be obtained in purified silicon at low temperature~\cite{Tyryshkin}. On the other hand, the coherence time can be improved through active quantum control of the many-body environment~\cite{Imamoglu2006,Yacoby2010,Togan_Nature2011,DreauPRL2014} or by decoupling the central spin from its fluctuations, either by applying periodic spin flips~\cite{Viola,Du,Hanson_Science2010} or by engineering spin eigenstates which are protected against environmental noise~\cite{Xu,Golter,Maletinsky_NatPhys2015,FuchsArxiv}. However, for these strategies to be effective, it is crucial to first identify the sources of noise and understand precisely their impact on the coherence properties of the central spin.\\
\indent Here we analyze how magnetic and electric field fluctuations impair the quantum coherence of the electronic spin associated with a single nitrogen-vacancy (NV) defect in diamond. This atomic-sized defect is attracting considerable interest for a broad range of applications including quantum metrology and sensing~\cite{Hodges,RondinRop,Schirhagl}, quantum information processing~\cite{MRS} and hybrid quantum systems~\cite{Zhu,Kubo,Arcizet}. For all these applications, optimal performances require a long spin coherence time. In this work, we analyze the contributions of magnetic and electric field fluctuations to spin decoherence by exploiting spin eigenstates protected either against magnetic noise or against electric noise~\cite{Dolde_NatPhys}. The competition between these noise sources is then analyzed quantitatively by changing their relative strength through modifications of the NV defect environment. \\
\indent The NV defect in diamond has a spin triplet ground state $S=1$ with a zero-field splitting $D\approx 2.88$~GHz between the $m_s=0$ and $m_s=\pm1$ spin sublevels, where $m_s$ denotes the spin projection along the NV symmetry axis $(z)$. The spin Hamiltonian describing the ground state in the presence of strain, electric field {\bf E} and magnetic field {\bf B} has been discussed in detail in Refs.~[\onlinecite{Dolde_NatPhys,Doherty_PRB2012}]. The strain, which is induced by a local deformation of the diamond crystal, can be treated as a local static electric field ${\bm \Sigma}$ interacting with the NV defect through the linear Stark effect~\cite{TamaratPRL2006}. Defining a total effective electric field $\bm \Pi= {\bm \Sigma} +{\bf E}$, the spin-Hamiltonian can be written as
\begin{multline}
H = (hD + d_{\parallel} \Pi_z) S_z^2 + g_e\mu_{\rm B} {\bf S}\cdot {\bf B} \\
\hspace{-3cm}- d_{\perp}\left[ \Pi_x(S_xS_y+S_yS_x)+\Pi_y(S_x^2-S_y^2)\right] \ ,
\label{hamiltonian}
\end{multline}
where ${\bf S}=\{S_x,S_y,S_z\}$ are the dimensionless electron spin operators, $h$ is the Planck constant, $d_{\parallel}/h=0.35$~Hz.cm.V$^{-1}$ and $d_{\perp}/h=17$~Hz.cm.V$^{-1}$ are the longitudinal and transverse components of the electric dipole moment~\cite{Vanoort}, $g_e$ is the electron $g$-factor and $\mu_{\rm B}$ is the Bohr magneton. For weak magnetic fields such that $B\ll hD/g_e\mu_{\rm B}$, the transverse components of the Zeeman interaction can be neglected and the eigenstates of the spin system are $\{\left|0\right.\rangle,\left|+\right.\rangle,\left|-\right.\rangle\}$, where
\begin{align*}
 & \left|+\right.\rangle=\cos\left(\frac{\theta}{2}\right)\left|+1 \right.\rangle+\sin\left(\frac{\theta}{2}\right)e^{i\phi}\left|-1 \right.\rangle \ , \\
& \left|-\right.\rangle=\sin\left(\frac{\theta}{2}\right)\left|+1 \right.\rangle-\cos\left(\frac{\theta}{2}\right)e^{i\phi}\left|-1 \right.\rangle \ .
\end{align*}
Here $\{\left|m_s \right.\rangle\}$ are the eigenstates of the $S_z$ operator, $\tan\phi=\Pi_{y}/\Pi_{x}$ and $$\tan\theta=\frac{\xi_{\perp}}{\beta_z} \ ,$$ where $\xi_{\perp}=d_{\perp}\sqrt{\Pi^2_{x}+\Pi^2_{y}}/h$ and $\beta_z=g_e\mu_{\rm B}B_z/h$. 

\begin{figure}[t]
\begin{centering}
\includegraphics[width=8.6cm]{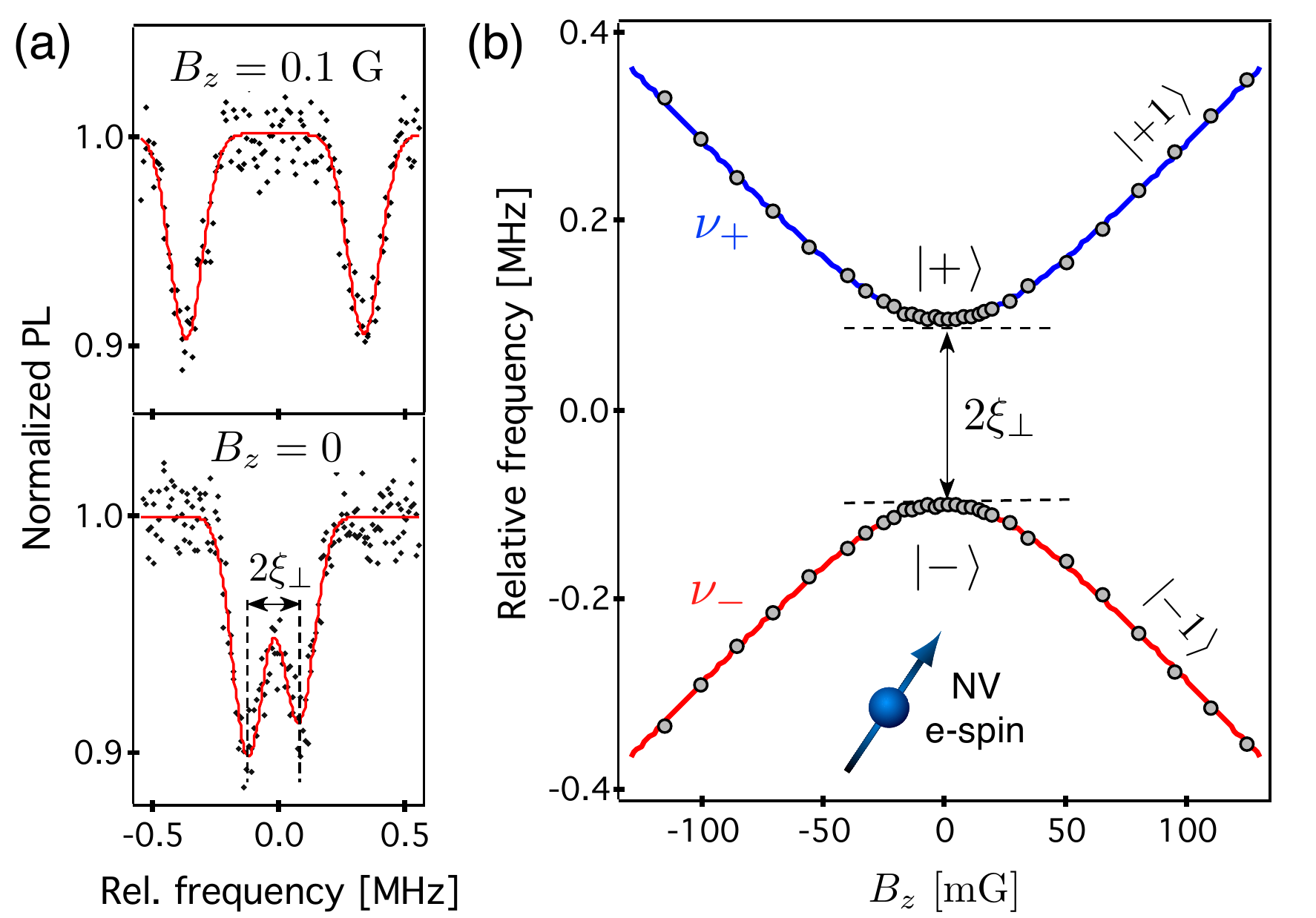}
\caption{(a)-Typical ESR spectra recorded from a single NV defect hosted in a high purity diamond crystal around $B_z=0$. The ESR transitions are associated with the $^{14}$N nuclear spin projection $m_I=0$. Those linked to $m_I=\pm1$ are not shown. (b)-ESR frequencies $\nu_+$ and $\nu_-$ as a function of $B_z$. The solid lines are data fitting with Eq. (2), leading to $\xi_{\perp}=93\pm2$~kHz.} 
\label{Fig1}
\end{centering}
\end{figure}

Since $d_{\parallel} \ll d_{\perp}$~\cite{Vanoort,sup}, we neglect the longitudinal component of the Stark effect and the frequencies $\nu_{\pm}$ of the electron spin resonance (ESR) transitions $\left|0\right.\rangle\rightarrow \left|\pm\right.\rangle$ are given by
 \begin{equation}
\nu_{\pm}= D \pm \sqrt{\xi_{\perp}^2+\beta_z^2} \ .
\end{equation}
In bulk diamond samples $\xi_{\perp}$ is in the range of $100$~kHz [Fig. 1] and can reach few MHz for NV defects hosted in nanodiamonds, where the intrinsic strain is much stronger~\cite{NeumannNJP2009}.

In the limit $\beta_z\gg \xi_{\perp}$, {\it i.e.} for $\theta \approx 0$, the eigenstates are those of the $S_z$ operator and the ESR frequencies evolve linearly with the axial magnetic field [see Fig.~1(b)]. In this regime, decoherence of the NV defect electron spin is usually dominated by magnetic field noise. On the other hand, if $\beta_z\ll \xi_{\perp}$, {\it i.e.} for $\theta \approx \pi/2$, the ESR frequencies are given by $\nu_{\pm}= D \pm \xi_{\perp}$ and the NV defect electron spin is protected against first-order magnetic field fluctuations since $\left \langle\pm\right| S_z \left|\pm\right.\rangle=0$. Decoherence is then dominated by strain/electric field noise and second-order (quadratic) magnetic field fluctuations. In the following, we analyze the impact of these fluctuations on the spin coherence by tuning the strength of $B_z$. In most of diamond samples, an enhancement of the coherence time is expected at zero magnetic field, as previously reported in Ref.~[\onlinecite{Dolde_NatPhys}].

\begin{figure*}[t]
\begin{centering}
\includegraphics[width=18.2cm]{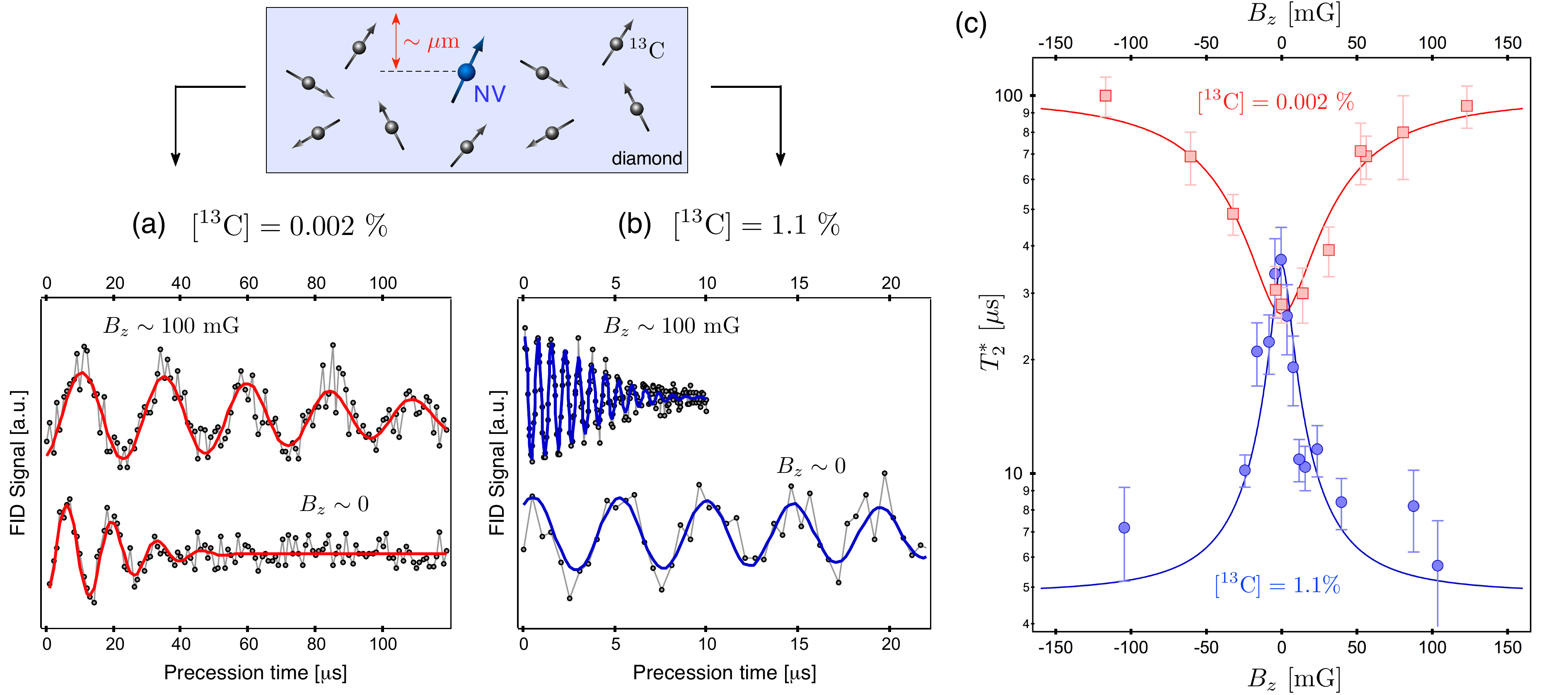}
\caption{(a,b)-Typical FID signals recorded at $B_z=0.1$~G (top) and $B_z\sim 0$ (bottom) from a native single NV defect hosted in a high-purity diamond sample with (a) $[^{13}{\rm C}]=0.002\%$ and (b) $[^{13}{\rm C}]=1.1\%$. In both cases, the NV defects are lying few micrometers below the diamond surface and $\xi_{\perp}\sim 100$~kHz. The coherence time $T_2^{*}$ is extracted by fitting the FID signal with the function $\cos(2\pi\Delta\tau)\times \exp{[-(\tau/T_2^{*})^2]}$, where $\Delta$ is the detuning between the microwave excitation and the ESR frequency. (c)-Evolution of $T_2^{*}$ (log scale) as a function of $B_z$ in the two samples with different content of $^{13}$C isotopes. The red solid line is data fitting with Eq.~(\ref{T2fit}). The blue solid line is a Lorentzian fit used as a guide-to-the-eye.}
\label{Fig1}
\end{centering}
\end{figure*}

Individual NV defects are optically isolated at room temperature using a scanning confocal microscope under green laser excitation. A coil is used to precisely control the magnetic field amplitude $B_z$ along the NV axis and ESR transitions are driven with a microwave field applied through a copper microwire spanned on the diamond surface. The nitrogen atom of the defect is a $^{14}$N isotope ($99.6 \%$ abundance), corresponding to a nuclear spin $I=1$. Each electron spin state is therefore split into three hyperfine sublevels. In the following, we focus on electron spin transitions associated with the $^{14}$N nuclear spin projection $m_I=0$, so that the spin Hamiltonian (1) is not modified by the hyperfine interaction~\cite{sup}. ESR spectra recorded from a single NV defect around zero magnetic field are shown in Fig.~1(a). The ESR frequencies closely follow Eq.~(2) with an anti-crossing at $B_z=0$, where $\nu_+-\nu_-=2\xi_{\perp}$ [Fig.~1(b)]. \\
\indent We first consider {\it native} NV defects hosted in an isotopically purified diamond crystal ($[^{13}\rm C]=0.002\%$) grown by chemical vapor deposition (CVD)~\cite{Teraji}. Decoherence of the NV defect electron spin is analyzed through measurements of the free-induction decay (FID) while applying the usual Ramsey sequence $(\pi/2)-\tau-(\pi/2)$~\cite{Maze2012}. Typical FID signals recorded at different magnetic field amplitudes $B_z$ are shown in Fig.~2(a). Surprisingly, a pronounced {\it dip} of the coherence time $T_2^{*}$ is observed around zero magnetic field [Fig.~2(c)]. \\
\indent To understand this behavior, we introduce the random variables $\delta \beta_z$ and $\delta \xi_{\perp}$, which describe the temporal fluctuations of the magnetic and electric fields around their mean values $\beta_z$ and $\xi_{\perp}$. In the limit $\beta_z\gg \xi_{\perp}$, the coherence time is given by $T_{2,\beta_z\gg \xi_{\perp}}^{*}=1/\sqrt{2}\pi\sigma_{\beta_z}$, where $\sigma^2_{\beta_{z}}=\left\langle \delta \beta_z^2\right.\rangle$ is the variance of the magnetic field fluctuations~\cite{sup}. In this regime, decoherence is governed by magnetic noise. When the spin system is approaching the level anti-crossing, the fluctuation of the ESR frequency $\delta \nu$ can be expressed as
\begin{equation*}
\delta\nu=\delta\xi_{\perp}\sin\theta + \delta \beta_z \cos\theta \ ,
\end{equation*}
leading to a decay of the FID signal with a coherence time~\cite{sup}
\begin{equation}
T_2^{*}=\frac{1}{\sqrt{2}\pi\sigma_{\xi_\perp}}\sqrt{\frac{1+\left(\frac{\beta_z}{\xi_\perp}\right)^2}{1+\left(\frac{\beta_z}{\xi_\perp}\right)^2\left(\frac{\sigma_{\beta_z}}{\sigma_{\xi_\perp}}\right)^2}} \ .
\label{T2fit}
\end{equation}
Here $\sigma^{2}_{\xi_{\perp}}=\left\langle \delta \xi_{\perp}^2\right.\rangle$ is the variance of the electric field fluctuations. We stress that such a simple analytic formula is valid (i) if $(\sigma_{\beta_z},\sigma_{\xi_{\perp}})\ll \xi_{\perp}$ and (ii) if the second-order magnetic field fluctuations of the ESR frequency can be neglected, {\it i.e.} for $\sigma^{2}_{\beta_{z}}/2\xi_{\perp}\ll \sigma_{\xi_{\perp}}$~\cite{sup}. 

At the level anti-crossing $T_{2,\beta_z=0}^{*}=1/\sqrt{2}\pi\sigma_{\xi_\perp}$, which indicates that the coherence time is limited by electric noise. In order to analyze the behavior of $T_2^*$ around the anti-crossing, we introduce the parameter 
\begin{equation}
\mathcal{R}=\frac{T_{2,\beta_z=0}^*}{T_{2,\beta_z\gg \xi_{\perp}}^*}=\frac{\sigma_{\beta_{z}}}{\sigma_{\xi_{\perp}}} \ .
\label{R}
\end{equation}
\indent If $\sigma_{\beta_{z}}<\sigma_{\xi_{\perp}}$, the coherence time drops around zero field ($\mathcal{R}<1$), as experimentally observed in Fig.~2(c) for a single NV defect hosted in an isotopically purified diamond sample. Data fitting with Eq.~(\ref{T2fit}) leads to $\sigma_{\beta_z}=2.20 \pm 0.06$~kHz and $\sigma_{\xi_{\perp}}=8.0\pm 0.2$~KHz, corresponding to $\mathcal{R}=0.28\pm 0.05$. This result reveals the existence of a significant source of electric field noise. It is known that a two-photon ionization process of the NV defect can promote charge carriers to the conduction band of diamond~\cite{Beha2012,Aslam2013,Siyushev2013}. This mechanism was recently used to demonstrate photoelectric detection of the electron spin resonance~\cite{Nesladek2015}. Here charge fluctuations induced by photo-ionization of the NV defect produce an electric field noise, which is likely the dominant decoherence mechanism in zero magnetic field. This source of electric noise is intrinsically linked to the optical illumination of the NV defect, which is required for polarization and readout of its electronic spin. For deep native NV defects in isotopically purified diamond samples, applying a static magnetic field enables to protect the central spin against this intrinsic source of electric noise.\\
\indent If $\sigma_{\beta_{z}}>\sigma_{\xi_{\perp}}$, the coherence time is expected to increase at the level anti-crossing. This regime can be reached by increasing $\sigma_{\beta_{z}}$. In high-purity diamond samples, the magnetic noise originates from the fluctuations of a bath of $^{13}$C nuclear spins ($I=1/2$). Increasing the amplitude of these fluctuations can be simply achieved by increasing the $^{13}$C content during the CVD growth~\cite{Balasu2009,Mizuochi}. Typical FID signals recorded around zero field for a single NV defect hosted in a commercial CVD-grown diamond sample with a natural content of $^{13}$C isotopes ($1.1\ \%$) are shown in Fig.~2(b). The evolution of $T_2^*$ with $B_z$ now reveals a coherence {\it peak} at zero magnetic field [Fig.~2(c)], as previously reported in Ref.~[\onlinecite{Dolde_NatPhys}]. This observation indicates that magnetic noise is now the strongest source of decoherence. In the limit $\beta_z\gg \xi_{\perp}$, we measure $T_{2,\beta_z\gg \xi_{\perp}}^*\sim 5 \ \mu$s corresponding to $\sigma_{\beta_z}\sim 40$~kHz. At the level anti-crossing, the static strain $\xi_{\perp}$ protects the central spin against first-order magnetic fluctuations leading to $T_{2,\beta_z=0}^*\sim 35 \ \mu$s, a value in the same range as the one obtained for single NV defects hosted in an isotopically purified diamond sample [Fig.~2(c)]. Here decoherence is fixed by the intrinsic electric field noise $\sigma_{\xi_\perp}$ and second-order magnetic field fluctuations ($\sigma^{2}_{\beta_{z}}/2\xi_{\perp}\sim 5$~kHz), which are reaching the same order of magnitude. In this regime, the simple model leading to Eq.~(\ref{T2fit}) is not valid and it is not possible to extract a simple analytic formula describing the full evolution of $T_2^*$ around the anti-crossing~\cite{sup}.\\
\indent We note that the linewidth of the coherence {\it peak} reaches $\Delta B_z\sim 10$~mG ($\Delta \beta_z\sim 30$~kHz). Such a narrow linewidth can be exploited to detect individual $^{13}$C nuclear spins weakly interacting with the NV defect through hyperfine coupling. This interaction can be modeled as an effective magnetic field, leading to a Zeeman shift of the ESR frequencies $\beta_h=\mathcal{A}_{\rm C}m_I$, where $\mathcal{A}_{\rm C}$ is the hyperfine coupling strength, which depends on the lattice site occupied by the $^{13}$C impurity~\cite{Smeltzer2011,Dreau2012}, and $m_I=\pm1/2$ is the nuclear spin projection along the NV axis. Level anti-crossings are then reached when $\beta_z+\beta_h=0$, {\it i.e.} for $\beta_z=\pm\mathcal{A}_{\rm C}/2$. Two coherence {\it peaks} can thus be observed around zero field, whose splitting is fixed by the hyperfine coupling strength [Fig.~3]. This method enables to detect weakly coupled $^{13}$C nuclei, {\it e.g.} $\mathcal{A}_{\rm C}\sim 50$~kHz in Fig.~3(b)].

\begin{figure}[t]
\begin{center}
\includegraphics[width=8.5cm]{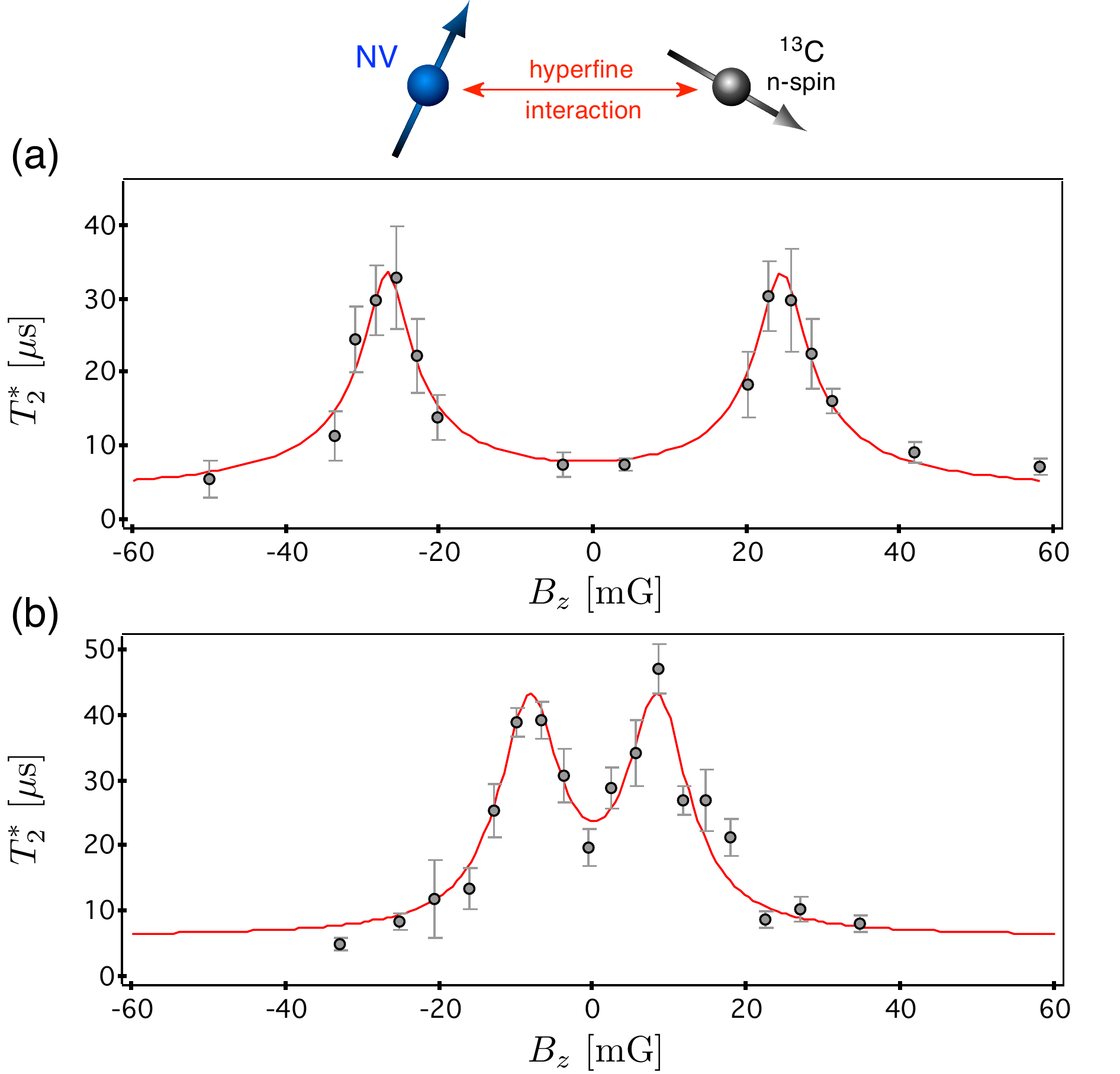}
\caption{$T_2^{*}$ as a function of $B_z$ for single NV defects coupled with a nearby $^{13}$C nuclear spin. The solid lines are data fitting with two Lorentzian functions. The hyperfine coupling strength extracted from the fit are (a) $\mathcal{A}_{\rm C}=144 \pm 2$~kHz and (b) $\mathcal{A}_{\rm C}=47 \pm 2$~kHz.}
\label{Fig3}
\end{center}
\end{figure}

\begin{figure}[t]
\includegraphics[width=7.5cm]{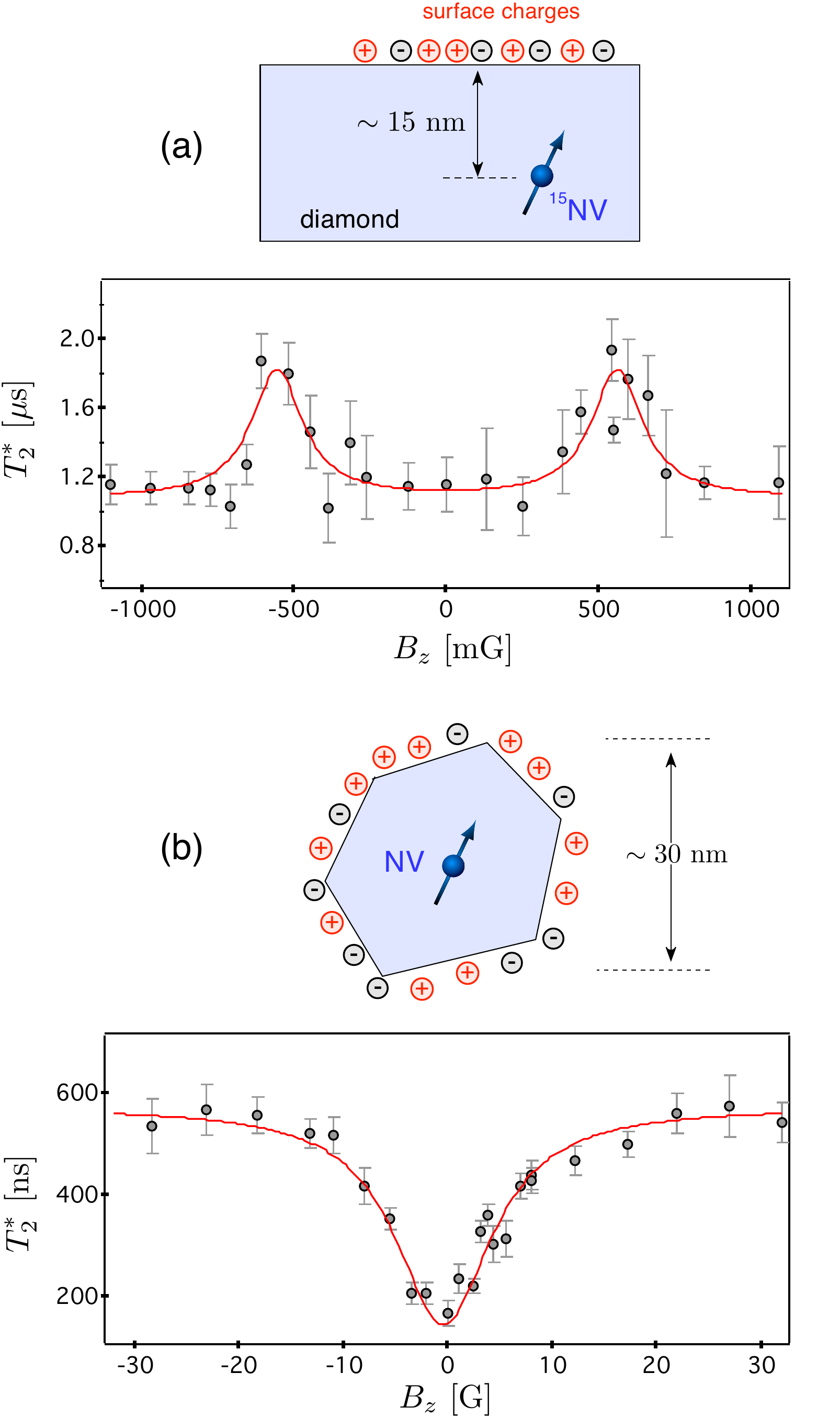}
\caption{(a) $T_2^{*}$ as a function of $B_z$ for a single NV defect implanted at $\sim 15$~nm below the surface of a high-purity diamond sample ($[^{13}\rm C]=1.1\%$). The static strain is $\xi_{\perp}=230$~kHz. (b) Same experiment realized for a single NV defect hosted in a 30-nm ND. The solid line is data fitting with Eq.~(\ref{T2fit}). Here the static strain is $\xi_{\perp}=7$~MHz.}
\label{Fig4}
\end{figure}

We now analyze how $T_2^*$ evolves around level anti-crossings while modifying the electric noise $\sigma_{\xi_{\perp}}$ surrounding the NV defect. To this end, we first investigate NV defects artificially created close to the surface of a high-purity diamond crystal ($[^{13}\rm C]=1.1\%$) through the implantation of $^{15}$N ions at $10$~keV. The diamond sample was then annealed during two hours in vacuum at $800^{\circ}$C, and its surface cleaned with acids. The resulting NV defects are located at roughly $15$~nm below the diamond surface and are associated with the $^{15}$N isotope, which is a $I=1/2$ nucleus characterized by a hyperfine coupling strength $\mathcal{A}_{^{15} \rm N}=3.15$~MHz~\cite{Rabeau2006}. A typical evolution of $T_2^{*}$ as a function of $B_z$ for a near-surface NV defect is shown in Fig.~4(a). Two coherence {\it peaks} are observed at $\beta_z=\pm \mathcal{A}_{^{15} \rm N}/2$, with a much smaller amplitude than the one observed for native NV defects placed few micrometers below the surface [Figs.~2(c) and 3]. For shallow-implanted defects, the electric field noise contribution $\sigma_{\xi_{\perp}}$ is expected to increase significantly, owing to the close vicinity of fluctuating charges lying on the diamond surface~\cite{Rugar2015}. Magnetic field fluctuations, which are also increased for near-surface NV defects~\cite{Degen2012}, remain however the strongest source of noise, resulting in an enhanced coherence time at the level-crossings~($\mathcal{R}>1$). \\
\indent For NV defects implanted closer to the surface ($\sim5$~nm) of an isotopically purified diamond sample, it was recently shown that decoherence could be dominated by electric field noise, even far away from the level anti-crossings~\cite{Rugar2015}. In this case, the longitudinal component of the Stark effect $\xi_{\parallel}=d_{\parallel}\Pi_z/h$ needs to be included in the model. Assuming that decoherence is dominated by electric field noise - regardless of the applied magnetic field - the parameter $\mathcal{R}$ defined by Eq.~(\ref{R}) becomes
$\mathcal{R}=\sigma_{\xi_{\parallel}}/\sigma_{\xi_{\perp}}=d_{\parallel}/d_{\perp}$, where $\sigma^2_{\xi_{\parallel}}$ is the variance of the electric field fluctuations along the $z$ axis~\cite{sup}. In this case, a huge {\it dip} of $T_2^*$ should therefore be observed around the level anti-crossing ($R\approx 1/50$). We note that in Ref.~[\onlinecite{Rugar2015}], the coherence time was inferred through dynamical decoupling sequence which are not sensitive to the same frequencies of the noise spectrum as Ramsey spectroscopy. Combining both approaches around level anti-crossings might be used in future to infer the spectral density of the electric field noise surrounding near-surface NV defects.\\
\indent In an attempt to access a regime with a larger contribution of the electric noise, we finally investigate single NV defects hosted in commercially available nanodiamonds (NDs) produced by milling type-Ib high-pressure high-temperature (HPHT) diamond crystals. The formation of NV defects was carried out using high energy ($13.6$~MeV) electron irradiation followed by annealing at $800^{\circ}$C under vacuum. The irradiated NDs were then oxidized in air at $550^{\circ}$C during two hours in order to remove graphitic-related defects on the surface and produce stable NV defects~\cite{Rondin2010}. The evolution of $T_2^{*}$ as a function of $B_z$ for a single NV defect hosted in a $30$-nm ND is shown in Fig.~4(b). A pronounced {\it dip} of the coherence time is observed at the level anti-crossing, which indicates that $\sigma_{\xi_{\perp}}>\sigma_{\beta_z}$ in NDs. This situation is similar to the one observed for native NV defects hosted in isotopically purified diamond samples [Fig. 2(c)]. Data fitting with Eq.~(\ref{T2fit}) leads to $\sigma_{\beta_z}=409 \pm 7$~kHz and $\sigma_{\xi_{\perp}}=1360\pm 50$~kHz. We note that the width of the {\it dip} is much larger than the one observed in bulk diamond samples, because the static strain reaches few MHz for NV defects hosted in NDs. \\
\indent Using Ramsey spectroscopy around a level anti-crossing, we have analyzed the competition between electric field and magnetic field fluctuations in the decoherence of the electronic spin associated with single NV defects in different types of diamond samples. To this end, we have used a static magnetic field to switch the spin system between eigenstates protected either against magnetic noise or against electric noise. This study provides significant insights into the decoherence of the NV electronic spin, thus giving new perspectives of performance optimization in quantum metrology and sensing applications~\cite{Hodges,RondinRop,Schirhagl}.\\

\noindent {\it Acknowledgements.} We thank R. B. Liu and P. Maletinsky for fruitful discussions and careful reading of the manuscript. This research has been supported by the European Union Seventh Framework Program (FP7/2007-2013) under the project DIADEMS and by the European Research Council  (ERC-StG-2014, IMAGINE). J.R.M acknowledges support from Conicyt-PIA grants No ACT1108 and ACT1112, Fondecyt grant No 1141185, and Air Force FA9550-15-1-0113.

\section*{Supplementary Information}
\subsection{Experimental setup}

Individual NV defects are optically isolated at room temperature using a home-built scanning confocal microscope under green laser excitation. Details of the experimental setup can be found in Ref.~[\onlinecite{Dreau2011}]. A coil is used to precisely control the magnetic field amplitude $B_z$ along the NV defect axis by canceling the Earth magnetic field ($\sim 0.5$ G). High-resolution ESR spectroscopy is achieved through repetitive excitation of the NV defect with a resonant microwave $\pi$-pulse followed by a $300$-ns read-out laser pulse~\cite{Dreau2011}. ESR spectra are recorded by continuously repeating this sequence while sweeping the $\pi$-pulse frequency and recording the PL intensity. Owing to spin-dependent PL of the NV defect, ESR is evidenced as a drop of the PL signal. The ESR spectra shown in Fig.1(a) of the main paper were obtained by setting the $\pi$-pulse duration to $6 \ \mu$s, as verified by recording electron spin Rabi oscillations.

For NV native defects in high-purity diamond samples, the $\pi/2$-pulse duration of the Ramsey sequence was set to $2 \ \mu$s, in order to selectively excite electron spin transitions associated with the $^{14}$N nuclear spin projection $m_I=0$. For NV defects hosted in nanodiamonds, the static strain $\xi_\perp$ is larger than the hyperfine interaction with the $^{14}$N nuclear spin, so that the MW pulses do not need to be selective.

\subsection{Random variable approach}

In this section, we introduce a simple model explaining the evolution of the NV defect electron spin coherence time $T_2^{*}$ around level anti-crossings. The notations are those introduced in the main text and we do not neglect the longitudinal component of the Stark effect $\xi_\parallel = d_\parallel\Pi_z/h$. 

In that case, the frequencies $\nu_{\pm}$ of the ESR transitions $\left|0\right.\rangle\rightarrow \left|\pm\right.\rangle$ are given by 
\begin{eqnarray}
\nu_\pm = D + \xi_\parallel \pm \sqrt{\xi_\perp^2 + \beta_z} \ .
\label{eq:energy}
\end{eqnarray}
In the following, we restrict the study to the $\{\ket{0}, \ket{+}\}$ spin subspace and we introduce the random variables $\{\delta \beta_z,\delta \xi_{\perp},\delta \xi_{\parallel}\}$, which describe the temporal fluctuations of the magnetic and electric fields around their central values $\{\beta_z,\xi_{\perp},\xi_{\parallel}\}$ with respective variances $\{\sigma_{\beta_z},\sigma_{\xi_\perp},\sigma_{\xi_\parallel} \}$.
The resulting fluctuation of the ESR frequency reads
\begin{eqnarray}
\delta\nu =  \delta\xi_\parallel + \sqrt{(\xi_\perp+\delta\xi_\perp)^2 + (\beta_z+\delta\beta_z)^2} - \sqrt{\xi_\perp^2 + \beta_z^2} 
\label{general}
\end{eqnarray}


\begin{table*}
\begin{center}
\begin{tabular}{|c|c|c|c|c|}

	\hline
    & [$^{13}$C]$=0.002 \%$ & [$^{13}$C]$=1.1 \%$ &  implanted NV & NV in nanodiamond\\
    & Fig. 2(a) and (c) & Fig. 2(b) and (c)  & Fig. 4(a) & Fig. 4(b)\\
	\hline
	$T_{2,\beta_z=0}^*$ & $28 \pm 3$  $\mu$s & $37 \pm 8$ $\mu$s & $1.88 \pm 0.16$ $\mu$s & $165 \pm 25$ ns \\
	\hline
	$T_{2,\beta_z\gg\xi_\perp}^*$ & $100 \pm 12$ $\mu$s & $5.4 \pm 1.5$ $\mu$s & $1.2 \pm 0.1$ $\mu$s & $550 \pm 40$ ns \\
	\hline
	$\xi_\perp$ & 50 kHz & 120 kHz & 230 kHz & 7 MHz\\
	\hline
	$\sigma_{\beta_z}$ & $2.2 \pm 0.06$ kHz & $41.5 \pm 2.6$ kHz & $187 \pm 3.5$ kHz & $409 \pm 7$ kHz\\
	\hline
	$\sigma_0$ & $8 \pm 0.2$ kHz & $6 \pm 0.3$ kHz & $120 \pm 2.3$ kHz & $1364 \pm 47$ kHz \\ 
	\hline
	$R$ & $0.28 \pm 0.05$ & $6.9 \pm 2.4$ & $1.5 \pm 0.2$ & $0.3 \pm 0.05$\\
	\hline
\end{tabular}
\end{center}
\caption{Values of the parameters extracted from the experiments presented in the main text.}
\label{Table1}
\end{table*}

Such a fluctuation induces dephasing of the NV center electronic spin, which is quantified through free-induction decay (FID) measurements. The spin system is polarized in state $\ket{0}$ before applying the usual Ramsey sequence $\frac{\pi}{2}-\tau - \frac{\pi}{2}$. The FID signal is the probability to retrieve the electronic spin in state $\ket{0}$ for a free precession time $\tau$, which is expressed as $$p_{\ket{0}} (\tau) = \left[ 1-\cos\right(\psi+\delta\psi\left)\right] /2 \ .$$ Here $\delta \psi = \int_0^\tau 2\pi \delta\nu dt$ and $\psi = \int_0^\tau 2\pi\Delta dt$, where $\Delta= \nu_+  - \nu_{\rm mw}$ is the detuning between the microwave excitation frequency $\nu_{\rm mw}$ and the ESR transition $\ket{0} \rightarrow \ket{+}$. After averaging, the result of the measurement is given by $\langle p_{\ket{0}} (\tau) \rangle = \left[ 1-\cos\psi\langle\cos\delta\psi\rangle + \sin\psi\langle\sin\delta\psi\rangle \right] /2$. Assuming that the random variable $\delta\psi$ is normally distributed, one can find 
\begin{eqnarray}
\langle p_0 (\tau) \rangle =\left[ 1-e^{-\langle\delta \psi^2\rangle/2} \cos\psi \right]/2 \ .
\label{eq:p0}
\end{eqnarray}
In order to predict the characteristic decay of the FID signal ($T_2^{*}$), one needs to extract the variance of the phase fluctuation 
\begin{eqnarray}
\langle \delta \psi^2 \rangle = 4\pi^2 \int_0^\tau dt \int_0^\tau dt'  \langle \delta \nu (t) \delta \nu (t') \rangle \ .
\label{eq:dpsi}
\end{eqnarray}

In the following, we consider random variables with correlation functions decaying exponentially
\begin{eqnarray}
\langle \delta \xi_\parallel (t) \delta \xi_\parallel (t') \rangle = \sigma_{\xi_{\parallel}}^{2} \exp(|t-t'|/\tau_{c,\xi_\parallel}) \ ,\\
\langle \delta \xi_\perp (t) \delta \xi_\perp (t') \rangle = \sigma_{\xi_{\perp}}^{2} \exp(|t-t'|/\tau_{c,\xi_\perp}) \ ,\\
\langle \delta \beta_z (t) \delta \beta_z (t') \rangle = \sigma_{\beta_{z}}^{2} \exp(|t-t'|/\tau_{c,\beta_z}) \ ,
\label{eq:correlation}
\end{eqnarray}
where $\{\tau_{c,\xi_\parallel},\tau_{c,\xi_\perp},\tau_{c,\beta_z}\}$ are the correlation times of electric and magnetic fluctuations. Furthermore, we assume that the magnetic and electric field noise sources are not correlated.

An analytic expression giving the characteristic decay of the FID signal as a function of $\beta_z$ can hardly be obtained with Eq.~(\ref{eq:dpsi}). To go further, we start by studying limit cases.

\subsubsection{Limit case ${\beta_z \gg \xi_{\perp}}$}
\label{Limit1}

We first consider the limit ${\beta_z \gg \xi_{\perp}}$. In this regime the ESR frequency evolves linearly with the magnetic field $\nu_{+}= D + \xi_\parallel + \beta_z$ and its characteristic fluctuation reads $$\delta\nu_{\beta_z \gg \xi_{\perp}}=\delta\xi_\parallel +\delta\beta_z \ .$$ 



For a slowly fluctuating bath~\cite{Taylor2008,Lange2010} such that $\tau \ll (\tau_{c,\xi_\parallel},\tau_{c, \xi_\perp}, \tau_{c,\beta_z})$, the variance of the resulting phase fluctuation is given by
\begin{eqnarray}
\langle \delta \psi^2 \rangle & = & 4\pi^2\left[ \sigma_{\xi_\parallel}^2 +\sigma_{\beta_z}^2 \right]\tau^2 \ ,
\label{eq:dpsislow}
\end{eqnarray}
leading to a characteristic decay of the FID signal 
\begin{eqnarray}
T_{2,\beta_z \gg \xi_{\perp}}^{*}=\frac{1}{\sqrt{2}\pi}\frac{1}{\sqrt{\sigma^2_{\beta_z}+\sigma^2_{\xi_{\parallel}}}} \ .
\label{eq:farLAC}
\end{eqnarray} 
If the longitudinal component of the Stark effect can be neglected, we obtain $T_{2,\beta_z \gg \xi_{\perp}}^{*}=1/\sqrt{2}\pi\sigma_{\beta_z}$, which is the formula used in the main text. In our experiments, the value of $\sigma_{\beta_z}$ was obtained by measuring the coherence time far from the level anti-crossing. Typical results obtained in the different samples studied in our work are summarized in Table S1.

\subsubsection{Limit case ${\beta_z \ll \xi_{\perp}}$}

In the limit ${\beta_z \ll \xi_{\perp}}$, the spin system is close to the level anti-crossing $\beta_z=0$. Assuming that $(\sigma_{\xi_\perp},\sigma_{\beta_z}) \ll \xi_{\perp}$, the fluctuation of the ESR frequency can then be expressed as $$\delta\nu_{\beta_z =0}\approx\delta\xi_\perp + \frac{\delta\beta_z^2}{2\xi_\perp} \ ,$$ where the last term describes the second-order (quadratic) magnetic field fluctuations. Here we have considered isotropic electric field fluctuations, so that $\sigma_{\xi_\parallel}\ll \sigma_{\xi_\perp}$ is always fulfilled. Following the same reasoning as in the previous paragraph, the coherence time is then given by 
\begin{eqnarray}
T_{2,\beta_z =0}^{*}= \frac{1}{\sqrt{2}\pi\sigma_0} \ ,
\label{eq:LAC}
\end{eqnarray} 
where $\sigma_0=\sqrt{\sigma^2_{\xi_{\perp}}+\sigma^4_{\beta_z}/4\xi_{\perp}^2}$. The values of $\sigma_0$ extracted through measurements of $T_2^*$ at zero magnetic field are summarized in Table S1.

\subsubsection{Parameter $\mathcal{R}$}

The parameter $\mathcal{R}$ is defined as the ratio between the coherence time obtained in the limit cases
\begin{equation}
\mathcal{R}=\frac{T_{2,\beta_z =0}^*}{T_{2,\beta_z \gg \xi_{\perp}}^{*}} =\sqrt{\frac{\sigma^2_{\beta_z}+\sigma^2_{\xi_{\parallel}}}{\sigma^2_{\xi_{\perp}}+\sigma^4_{\beta_z}/4\xi_{\perp}^2}} \ .
\end{equation}
Several behavior of the coherence time can be obtained around the level anti-crossing depending on the relative amplitudes of the different types of fluctuations.
\begin{itemize}
\item If decoherence is dominated by electric field fluctuations - regardless of the applied magnetic field~\cite{Rugar2015}- $\mathcal{R}=\sigma_{\xi_{\parallel}}/\sigma_{\xi_{\perp}}=d_{\parallel}/d_\perp\approx 1/50$. A huge drop of the coherence time should therefore be observed at the level anti-crossing.
\item Conversely, if decoherence is dominated by magnetic field fluctuations - even at $\beta_z= 0$ - then $\mathcal{R}=2 \xi_{\perp}/\sigma_{\beta_z}$ and a coherence peak is obtained at the anti-crossing. In this case the enhancement of the coherence time could be modified by tuning the static strain $\xi_{\perp}$ at the NV defect location, {\it e.g.} using a tip pressing on the diamond surface~\cite{MaletinskyPRL}.
\item If second-order magnetic fluctuations can be neglected then $\mathcal{R}=\sigma_{\beta_z}/\sigma_{\xi_{\perp}}$. In this case, a simple analytical formula describing the evolution of $T_2^{*}$ around the anti-crossing can be obtained [see Section~\ref{Analytic}]. In our work, this regime is obtained (i) for native NV defects in an isotopically purified diamond sample [Fig. 2(a)] and (ii) for NV defects hosted in diamond nanocrystals [Fig. 4(b)]. In both cases, a dip of the coherence time was observed, which indicates that $\sigma_{\xi_{\perp}}>\sigma_{\beta_z}$.
\item In intermediate regimes, the electric field fluctuations $\sigma_{\xi_{\perp}}$ and the second-order magnetic fluctuations are competing in zero magnetic field. This is the experimental situation obtained (i) for native NV defects in a high-purity diamond sample with a natural content of $^{13}$C isotope [Figs. 2(b) and 3 of the main text] and (ii) for shallow-implanted NV defects [Fig. 4(a) of the main text]. 
\end{itemize}

\subsubsection{Linear fluctuation regime}
\label{Analytic}

When second-order magnetic field fluctuations can be neglected at the anti-crossing ($\sigma^{2}_{\beta_{z}}/2\xi_{\perp}\ll \sigma_{\xi_{\perp}}$), the fluctuation of the ESR frequency can be expressed as 
\begin{eqnarray}
\delta \nu = \delta \xi_\parallel + \delta\xi_\perp\sin\theta +\delta\beta_z \cos\theta \ , 
\label{eq:dnu}
\end{eqnarray}
where $\tan\theta=\frac{\xi_{\perp}}{\beta_z}$. Using the same reasoning as in Section~\ref{Limit1} and neglecting $\delta \xi_\parallel $, the coherence time $T_2^*$ is given by
\begin{eqnarray}
T_2^*=\frac{1}{\sqrt{2}\pi \sigma_{\xi_\perp}}\sqrt{ \frac{1+\left(\frac{\beta_z}{\xi_\perp}\right)^2}{1+\left(\frac{\beta_z}{\xi_\perp}\right)^2\left(\frac{\sigma_{\beta_z}}{\sigma_{\xi_\perp}}\right)^2} }.
\label{eq:T2}
\end{eqnarray}
corresponding to Eq.~(3) of the main text. This formula was used to fit the experimental results obtained (i) for native NV defects in an isotopically purified diamond sample [Fig. 2(c)] and (ii) for NV defects hosted in diamond nanocrystals [Fig. 4(b)].

\end{document}